\documentclass[aps,prl,twocolumn,superscriptaddress,groupedaddress, amsmath,amssymb]{revtex4-2}

\usepackage{graphicx} 
\usepackage{physics} 
\usepackage{bm} 
\usepackage{hyperref}
\usepackage{xcolor}

\hypersetup{ colorlinks=true, linkcolor=blue, filecolor=magenta, urlcolor=blue, citecolor=blue }

\begin{document}


\title{Twist-Tunable Paramagnetic Superconductivity in $d$-wave Altermagnet/Superconductor Heterostructures}

\author{Narges Kia} \affiliation{Department of Physics, Institute for Advanced
Studies in Basic Sciences (IASBS), Zanjan 45137-66731, Iran}

\author{Saeed H. Abedinpour} \affiliation{Department of Physics, Institute for
Advanced Studies in Basic Sciences (IASBS), Zanjan 45137-66731, Iran}

\author{Zahra Faraei} 
\email{z.faraei@iasbs.ac.ir}
\affiliation{Department of Physics, Institute for Advanced
Studies in Basic Sciences (IASBS), Zanjan 45137-66731, Iran}

\date{\today}

\begin{abstract} 
The interplay between twist-angle engineering and unconventional magnetism provides a powerful new route to control quantum phenomena. We theoretically investigate a heterostructure comprising a $d$-wave superconductor proximitized by a two-dimensional $d$-wave altermagnet. We reveal that the momentum-space mismatch between the superconducting gap nodes and the altermagnetic spin-splitting nodes generates a robust, twist-tunable odd-frequency spin-triplet pairing. Consequently, the macroscopic electromagnetic response of the system can be tuned from a conventional diamagnetic Meissner state to an anomalous paramagnetic Meissner effect driven entirely by the interfacial twist angle. For a $d_{x^2-y^2}$ altermagnet, the paramagnetic response is maximized at perfect alignment ($\phi=0$) and completely suppressed at a maximal twist of $\phi=\pi/4$, while a $d_{xy}$ altermagnet exhibits the exact complementary behavior. Our results establish twisted altermagnetic heterostructures as a versatile platform for engineering odd-frequency pairing and macroscopic superconducting phases.
\end{abstract}

\maketitle

\textit{Introduction---}Altermagnets (AMs) have recently emerged as a unique class of collinear magnetic materials~\cite{Ahn2019a,Smejkel_SciAdv2020,Mazin2022, Smejkal2022}, combining the compensated macroscopic magnetization of antiferromagnets with the momentum-dependent spin splitting typical of ferromagnets \cite{Smejkal2022symmetry, Smejkal2022_PRX, Jungwirth_Newton2025, Ling_AdvFuncMat2024,Song_NatRevMat2025}. 
This spin degeneracy lifting is driven by crystal symmetries rather than relativistic effects, yielding distinct $d$-, $g$-, or $i$-wave spin-splitting textures. 
AMs offer an intriguing building block for superconducting heterostructures~\cite{Fukaya2025, Maeda_PRB2025, Zhao_PRB2025, Ouassou2023, Alam2026, Heinsdorf2026, Vakili2026,Fukaya_JPCM2025, liu_arxiv2025}. They allow for the magnetic control of superconducting states without introducing the destructive stray fields of ferromagnets or facing the spin-degeneracy limitations of conventional antiferromagnets.

When an even-frequency spin-singlet superconductor (SC) is placed in proximity to a spin-dependent environment, interfacial scattering can generate spin-triplet odd-frequency Cooper pairs \cite{Berezinskii1974,Balatsky_PRB1992,Bergeret_RMP2005,Linder2015}. In this unconventional pairing state, the anomalous pair amplitude is antisymmetric in relative time, requiring a compensating change in spin or spatial symmetry to satisfy the Pauli principle \cite{Bergeret_RMP2005,Linder2015,Tanaka2012}. Beyond its fundamental interest, odd-frequency pairing is known to generate anomalous electromagnetic responses, most notably the paramagnetic Meissner effect (PME) \cite{DiBernardo_PRX2015,DiBernardo_NatureComm2015,Asano2011}. However, observing and controlling the PME in solid-state devices has proven exceedingly difficult. Most proposals so far rely on ferromagnet/superconductor interfaces, where the strong exchange fields needed for triplet conversion also cause severe orbital and Pauli depairing. This effectively suppresses the parent superconducting state and obscures the PME.

\begin{figure}[t] \centering
\includegraphics[width=0.9\linewidth]{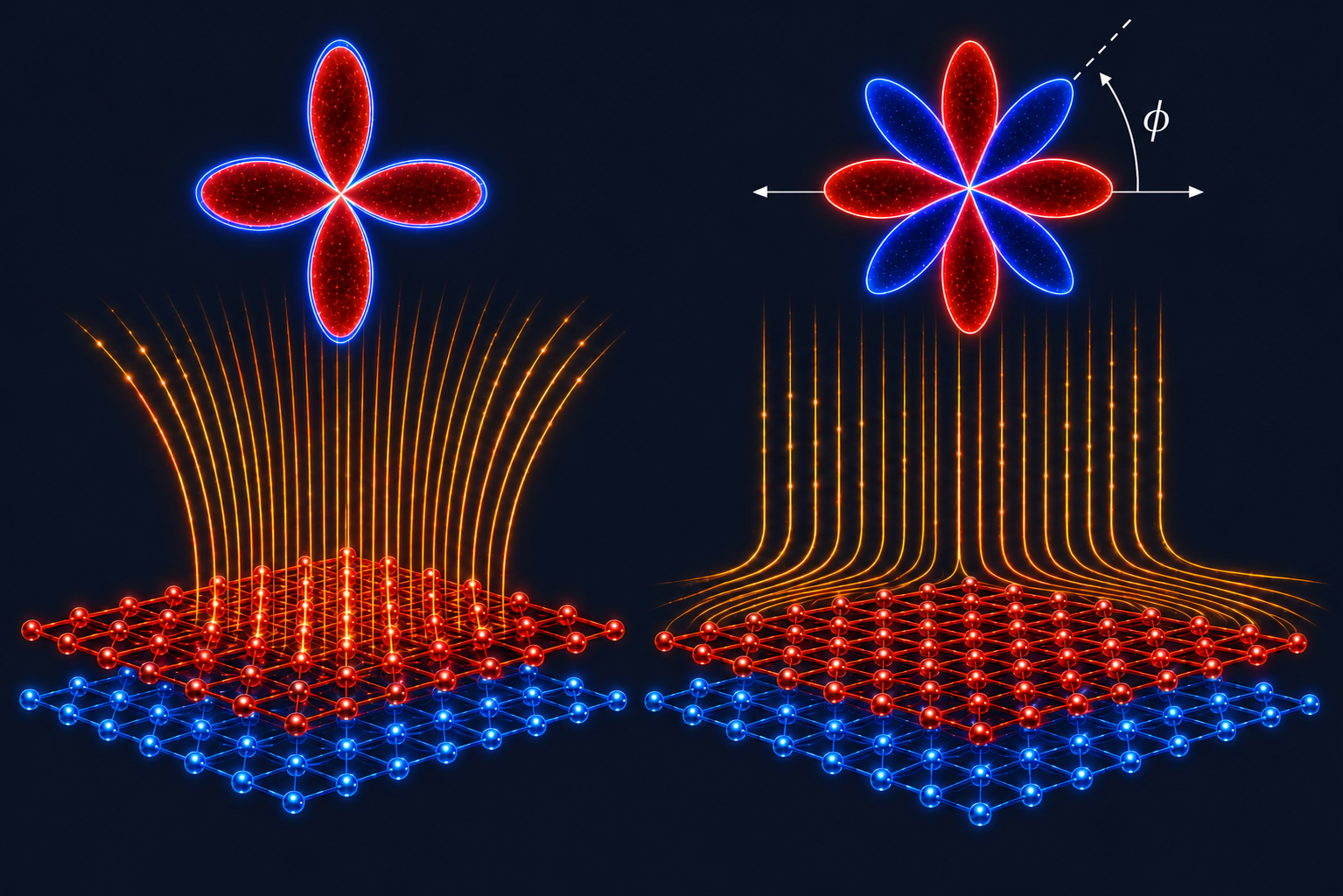}
\caption{Twist-angle tuning of the electromagnetic response in a $d$-wave altermagnet/superconductor heterostructure.
Perfect alignment of the $d$-wave nodes ($\phi = 0$, left) drives robust odd-frequency triplet pairing, leading to a paramagnetic Meissner state (flux attraction). Conversely, an interfacial twist ($\phi \neq 0$, right) induces a nodal mismatch that suppresses the paramagnetic response and moves the system toward a conventional diamagnetic state (flux expulsion).} \label{fig:schematic}
\end{figure}

In this Letter, we propose a route to achieve and control paramagnetic superconductivity using twisted heterostructures comprising a $d$-wave altermagnet and a $d$-wave superconductor (Fig.~\ref{fig:schematic}). Such devices could be realized by mechanically assembling cuprates with emerging van der Waals (vdW) altermagnets~\cite{Jiang2025, Sodequist2024,Cui2024}. Because AMs are collinear and macroscopically compensated, they induce spin-dependent scattering—yielding $S_z=0$ spin-triplet pairs~\cite{Chakraborty_PRB2025}—without the destructive orbital depairing inherent to ferromagnets. We show that the interfacial twist angle $\phi$ acts as a direct tuning knob for the electromagnetic response by modifying the momentum-space overlap between the superconducting gap and the AM spin-splitting texture. Specifically, when a $d_{x^2-y^2}$ AM is aligned with a $d$-wave SC, the coinciding antinodes maximize the generation of odd-frequency, extended-$s$-wave triplet correlations, overcoming local Pauli depairing to push the system into a robust paramagnetic superconducting regime. Rotating the interface by $\phi = \pi/4$ entirely eliminates this overlap, fully suppressing the odd-frequency conversion and recovering a standard diamagnetic response. We also find complementary behavior in $d_{xy}$ altermagnets, where the phase boundaries simply shift by $45^\circ$. 

\textit{Model and Formulation---}We consider a two-dimensional heterostructure, illustrated in Fig.~\ref{fig:schematic}, consisting of a $d$-wave altermagnet monolayer coupled to a pristine $d$-wave superconductor. Assuming a square lattice, the two-band normal-state Hamiltonian of the AM layer is described by $\hat{\mathcal{H}}_{AM}(\mathbf{k}) = \xi_{\mathbf{k}}\sigma_0 + h_{\mathbf{k}}\sigma_z$, adopting the standard phenomenological model for altermagnets~\cite{Smejkal2022_PRX}, where $\sigma_0$ and $\sigma_z$ are the Pauli spin matrices. While our theoretical framework seamlessly accommodates any $d$-wave symmetry, we focus our main text analysis on the $d_{x^2-y^2}$ altermagnet for pedagogical clarity, reserving the complementary $d_{xy}$ case for the End Matter. 
The kinetic energy and the $d_{x^2-y^2}$ altermagnetic exchange field are given by
\begin{align} 
\xi_{\mathbf{k}} &= -2t(\cos k_x + \cos k_y) - \mu, \nonumber \\ 
h_{\mathbf{k}} &= 2J(\cos k_x - \cos k_y). 
\end{align}
Here, $t$ is the nearest neighbor hopping integral, $\mu$ is the chemical potential, and $J$ dictates the strength of the $d$-wave spin-splitting.

The pristine SC layer hosts a $d_{x^2-y^2}$-wave gap. Assuming the standard limit of specular tunneling governed by momentum-conserving processes at the interface~\cite{Pixley_AnnRev2026}, the SC gap projected onto the AM layer becomes a function of the rotated momentum coordinates $(k'_x,k'_y)$ due to a relative twist angle $\phi$ between the layers. These are explicitly related to the unrotated coordinates $(k_x,k_y)$ via:
\begin{align}
  &k'_x=k_x \cos\phi - k_y \sin\phi, \nonumber
  \\
&k'_y=k_x\sin\phi+k_y\cos\phi.  
\end{align}
Consequently, the effective proximitized SC gap takes the form $\Delta_\phi(\mathbf{k}) =\Delta_d \Big[ \cos(k'_x) - \cos(k'_y) \Big]$, where $\Delta_d$ is the maximum gap amplitude.

Assuming a bulk SC layer, we neglect the inverse proximity effect. Integrating out the SC degrees of freedom via a phenomenological tunneling Hamiltonian yields a retarded self-energy in the AM layer. Extending the Green's function proximity formalism recently introduced for conventional $s$-wave SC/AM junctions~\cite{Alam2026} to our twisted $d$-wave architecture, we find that this proximity effect dresses the AM electrons~\cite{supplemental}. This gives rise to a renormalized Matsubara frequency $\tilde{\omega}_n$ and a highly momentum-dependent induced pairing potential $\tilde{\Delta}_n$: 
\begin{align} 
\tilde{\omega}_n(\mathbf{k}, \phi) &= \omega_n \left( 1 + \frac{\lambda_s}{\sqrt{\Delta_\phi^2(\mathbf{k}) + \omega_n^2}} \right), \nonumber \\ 
\tilde{\Delta}_n(\mathbf{k}, \phi) &= \frac{\lambda_s \Delta_\phi(\mathbf{k})}{\sqrt{\Delta_\phi^2(\mathbf{k}) + \omega_n^2}},
\end{align} 
where $\lambda_s$ denotes the effective proximity coupling strength.

\begin{figure}[t] \centering
\includegraphics[width=\linewidth]{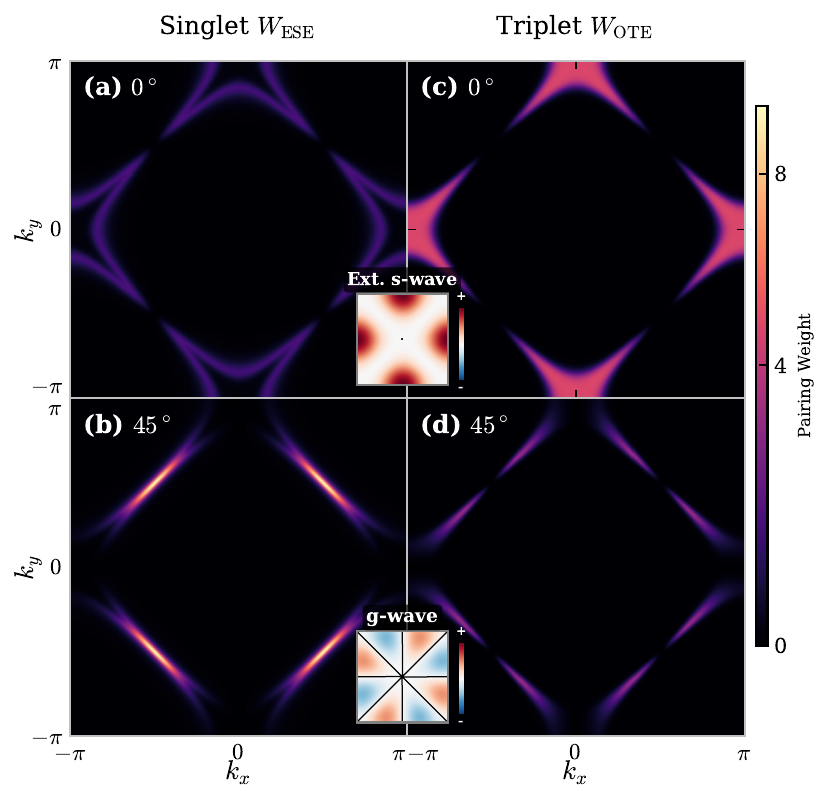} 
\caption{Numerical heatmaps of the pairing weights $W_{\rm ESE}$ [(a),(b)] and $W_{\rm OTE}$ [(c),(d)] for the perfectly aligned ($\phi=0^\circ$) [(a),(c)] and maximally twisted ($\phi=45^\circ$) [(b),(d)] cases. The insets show the density plots of the geometrical overlap product $h_{\mathbf{k}} \Delta_\phi(\mathbf{k})$ in the altermagnet's Brillouin zone for $\phi=0^\circ$ (top) and $\phi=45^\circ$ (bottom), illustrating the extended-$s$-wave and $g$-wave symmetries, respectively. Model parameters are set to $\mu=0$, $J=0.05t$, $\lambda_s=0.15t$, $\Delta_d=0.03t$, and $k_{\rm B}T=0.02t$. }
\label{fig:kspace} 
\end{figure}

\textit{Odd-Frequency Triplet Generation---}The coexistence of the induced pairing gap $\tilde{\Delta}_n$ and the altermagnetic field $h_{\mathbf{k}}$ dynamically generates spin-triplet correlations. By analytically inverting the dressed Nambu-Gor'kov matrix, we extract the anomalous pairing component $\hat{F}(i\omega_n, \mathbf{k}, \phi)$ of the AM layer (see Supplemental Material~\cite{supplemental} for details). The broken spin-rotation symmetry splits $\hat{F}$ into a spin-singlet even-frequency component $\psi$ and a mixed-spin ($S_z=0$) spin-triplet odd-frequency component $d_z$. 

Defining the characteristic denominator $\chi_n(\mathbf{k}, \phi) = h_{\mathbf{k}}^2 - \xi_{\mathbf{k}}^2 - \tilde{\omega}_n^2 - \tilde{\Delta}_n^2(\mathbf{k}, \phi)$, the exact analytical forms of the induced spin-singlet and spin-triplet amplitudes evaluate to:
\begin{subequations}\label{eq:pairing_amplitudes}
\begin{align} 
\psi(i\omega_n, \mathbf{k}, \phi) &= \frac{\chi_n(\mathbf{k}, \phi) \tilde{\Delta}_n(\mathbf{k}, \phi)}{\chi_n^2(\mathbf{k}, \phi) + 4h_{\mathbf{k}}^2 \tilde{\omega}_n^2},
\label{eq:psi}
\\
d_z(i\omega_n, \mathbf{k}, \phi) & = \frac{2 i \tilde{\omega}_n h_{\mathbf{k}}  \tilde{\Delta}_n(\mathbf{k}, \phi)}{\chi_n^2(\mathbf{k}, \phi) + 4 h_{\mathbf{k}}^2 \tilde{\omega}_n^2}. 
\label{eq:dz}
\end{align}
\end{subequations}
The contrasting symmetries of these pairing channels are immediately encoded in their respective numerators. Since the characteristic denominator and the effective gap $\tilde{\Delta}_n$ are strictly even in Matsubara frequency, the singlet amplitude $\psi$ adheres to the conventional even-frequency, spin-singlet, even-parity (ESE) symmetry class. In striking contrast, the presence of the purely imaginary, frequency-odd factor $i\tilde{\omega}_n$ in Eq.~\eqref{eq:dz} rigorously dictates that the dynamically generated spin-triplet correlations belong to the exotic odd-frequency, spin-triplet, even-parity (OTE) symmetry class. Crucially, because $\tilde{\omega}_n$ preserves spatial symmetries, the macroscopic momentum-space profile and the robustness of this anomalous odd-frequency state are fundamentally governed by the geometrical overlap product $h_{\mathbf{k}} \Delta_\phi(\mathbf{k})$, illustrated in the insets of Fig.~\ref{fig:kspace}.

To further quantify ESE and OTE pairings, we define the momentum-resolved pairing weights $W_{\rm ESE}(\mathbf{k}) = \sum_n |\psi|^2$ for the ESE and $W_{\rm OTE}(\mathbf{k}) = \sum_n |d_z|^2$ for the OTE pairings. Figs.~\ref{fig:kspace}(a-d) illustrate our numerical evaluation of these weights for untwisted and maximally twisted structures. For the fully aligned heterostructure ($\phi=0$) [see Fig.~\ref{fig:schematic} (left)], the antinodes of the AM and SC layers perfectly align. The product $h_\mathbf{k}\Delta_\phi(\mathbf{k})\propto(\cos{k_x}-\cos{k_y})^2$ is positive-definite, implying that the induced odd-frequency term does not change sign under a $\pi/2$ rotation, thus exhibiting a robust extended-$s$-wave symmetry [Fig.~\ref{fig:kspace}, panel (c) and upper inset].

 In contrast, at the maximum twist angle ($\phi=\pi/4$) [see Fig.~\ref{fig:schematic} (right)], the rotated $d$-wave SC gap effectively assumes a $d_{xy}$ symmetry. The antinodes of $\Delta_{\pi/4}$ strictly align with the nodes of $h_{\mathbf{k}}$, causing the geometrical overlap product $h_{\mathbf{k}} \Delta_\phi(\mathbf{k})$ to assume a $g$-wave pattern [lower inst of Fig.~\ref{fig:kspace}]. Because the maxima of the SC gap now coincide with the regions of zero altermagnetic field, this spatial mismatch severely attenuates the local triplet conversion everywhere in the Brillouin zone [Fig.~\ref{fig:kspace}(d)]. Consequently, the local triplet pairing weight $W_{\rm OTE}(\mathbf{k})$ is heavily suppressed, setting the stage for the unperturbed singlet component $W_{\rm ESE}(\mathbf{k})$ to dominate the system's macroscopic properties.

\textit{Twist-Tunable Paramagnetic Meissner Effect---}The most profound macroscopic consequence of odd-frequency superconductivity is a diamagnetic-to-paramagnetic transition in the electromagnetic response \cite{DiBernardo_PRX2015,DiBernardo_NatureComm2015, Asano2011, Fominov2015, Yokoyama2011}. To determine the global magnetic phase, we calculate the total normalized superfluid density $\rho_s(\phi)$ within linear response theory \cite{Asano2011, Fominov2015}:
\begin{equation} 
\rho_s(\phi) \propto T \sum_{\omega_n} \int_{\text{BZ}} d^2k \ \mathcal{W}(\mathbf{k}) \Big[ |\psi|^2 - |d_z|^2 \Big], 
\end{equation} 
where $T$ is the temperature and $\mathcal{W}(\mathbf{k}) =\sum_{i=x,y} (v_i^2 - u_i^2)$ is the effective kinetic weight factor governed by the competition between the normal Fermi velocity ($v_i = \partial \xi_{\mathbf{k}} / \partial k_i$) and the spin-dependent altermagnetic velocity ($u_i = \partial h_{\mathbf{k}} / \partial k_i$). For a $d_{x^2-y^2}$ altermagnet, this explicitly evaluates to $\mathcal{W}(\mathbf{k}) = 4(t^2 - J^2)(\sin^2 k_x + \sin^2 k_y)$. The negative sign preceding $|d_z|^2$ is the macroscopic hallmark of odd-frequency pairing, originating from its anomalous paramagnetic current response \cite{Fominov2015, Yokoyama2011, Linder2015}. A negative total superfluid density ($\rho_s < 0$) signifies the emergence of the paramagnetic Meissner effect \cite{Suzuki2014, Mironov2012}. Notably, as the spin splitting approaches the bandwidth limit ($J \to t$), not only the weight factor $\mathcal{W}(\mathbf{k})$ analytically vanishes, both spin-singlet and spin-triplet pairing amplitudes also become infinitesimally negligible, effectively suppressing the superfluid response and any induced superconductivity in the AM layer.

\begin{figure}[t] \centering
\includegraphics[width=\linewidth]{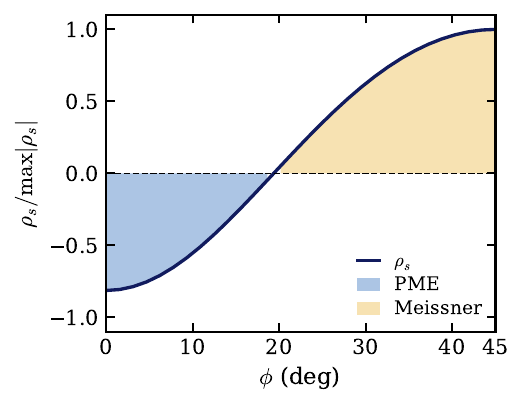} 
\caption{Normalized total superfluid density $\rho_s$ versus the twist angle $\phi$ for a $d_{x^2-y^2}$ altermagnet. The structural rotation drives the system from a paramagnetic Meissner state ($\rho_s < 0$) into a conventional diamagnetic state ($\rho_s > 0$) across a critical angle $\phi_c \approx 20^\circ$. Parameters chosen here: $J = 0.1 t$, $\mu=0$, $\lambda_s=0.15 t$, $\Delta_d=0.3 t$, and $k_{\rm B}T = 0.02 t$.} 
\label{fig:rhos} \end{figure}

As shown in Fig.~\ref{fig:rhos}, for $\phi \approx 0$, the dominant extended-$s$-wave odd-frequency component overpowers the singlet contribution in the kinetic integral, yielding a robust paramagnetic state ($\rho_s < 0$). However, as the twist angle increases, the symmetry mismatch shifts this delicate balance. At a critical twist angle $\phi_c \approx 20^\circ$, the macroscopic singlet and triplet kinetic contributions perfectly cancel each other ($\rho_s = 0$). For $\phi > \phi_c$, the severe spatial mismatch heavily suppresses the magnitude of the triplet pairing $|d_z|^2$. Consequently, the singlet component $W_{\rm ESE}$ reclaims macroscopic dominance over the kinetic integral, restoring the conventional even-frequency diamagnetic response ($\rho_s > 0$) despite a surviving, albeit highly attenuated, local triplet fraction.

\begin{figure}[t] \centering
\includegraphics[width=\linewidth]{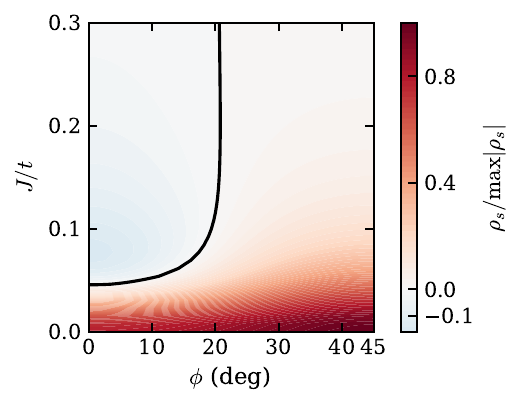}
\caption{Magnetic phase diagram of the twisted $d_{x^2-y^2}$-wave AM/$d$-wave SC heterostructure in the parameter space of exchange strength $J$ and twist angle $\phi$. The paramagnetic Meissner effect (blue region) forms a highly localized pocket, requiring both sufficiently strong exchange fields to generate triplet pairs and small angular misalignments to preserve momentum-space overlap. The other model parameters are $\mu=0$, $k_{\rm B}T=0.02 t$, $\Delta_d=0.03 t$, and $\lambda_s=0.15 t$.}
\label{fig:phasediagram} \end{figure}

This behavior is comprehensively mapped in the global ($J, \phi$) phase diagram (Fig.~\ref{fig:phasediagram}). The paramagnetic superconducting phase acts as a localized pocket. It manifests exclusively when the altermagnetic spin splitting is sufficiently strong to drive the singlet-to-triplet conversion, and the twist angle is small enough to maintain the requisite geometric alignment between the two strongly anisotropic momentum-space $d$-wave profiles.

\begin{figure}[t] \centering
\includegraphics[width = \linewidth]{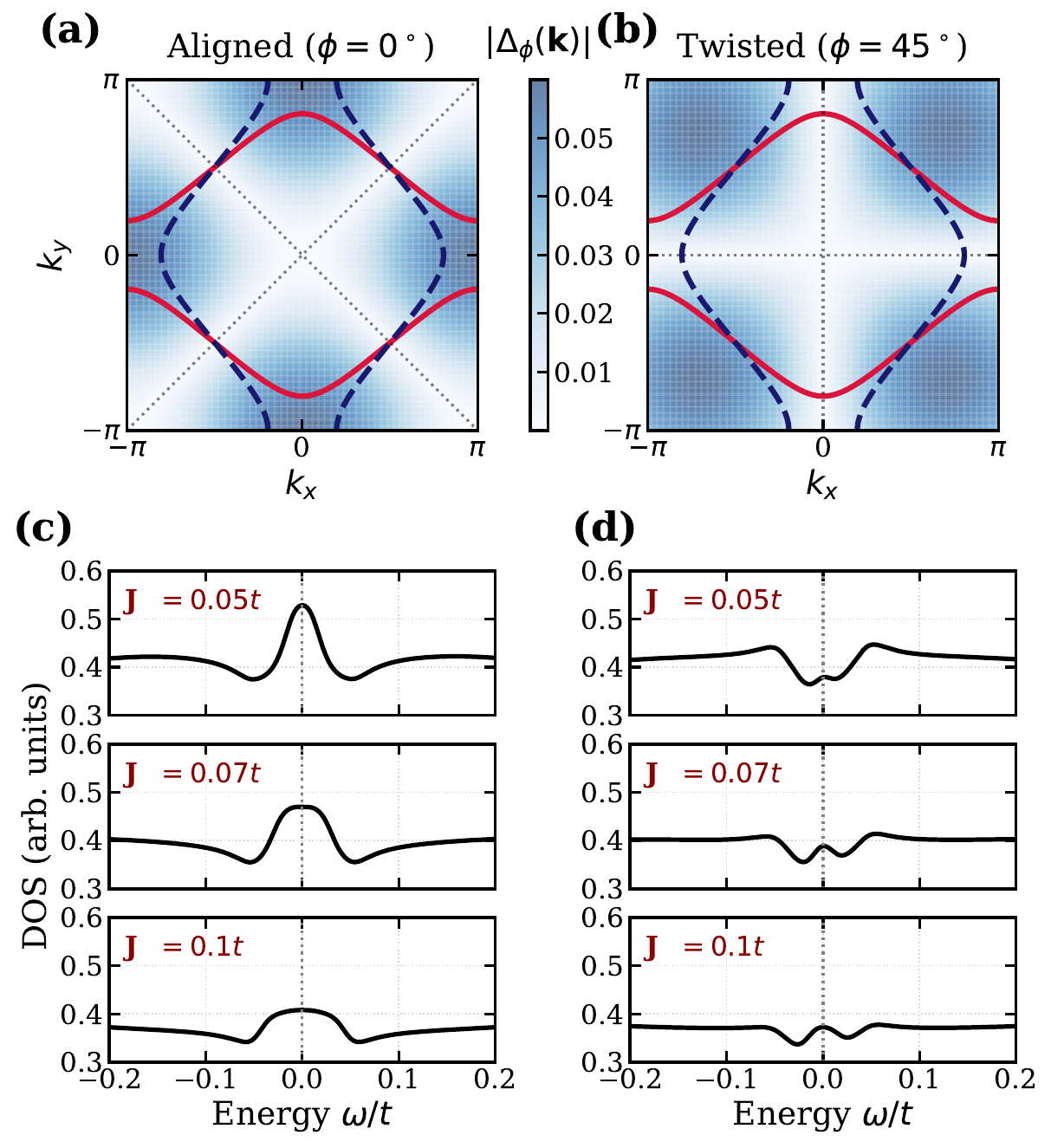} 
\caption{Direct correlation between momentum-space alignment and the emergence of odd-frequency resonances at half-filling ($\mu=0$). (a),(b) Spin-split Fermi surfaces overlaid on the magnitude of the underlying superconducting gap $|\Delta_\phi(\mathbf{k})|$ (blue shading) for  $J=0.1t$. (c),(d) Evolution of the corresponding total density of states (DOS) for varying exchange strengths $J$. The other model parameters are $\lambda_s=0.15t$, $\Delta_d=0.03 t$, and $\Gamma=0.5\Delta_d$.}
\label{fig:ldos} \end{figure}

\textit{Density of States---}A direct single-particle observable for the emergence of odd-frequency pairing is the appearance of low-energy subgap resonances in the density of states (DOS), which is directly proportional to the differential conductance $dI/dV$ measured in scanning tunneling spectroscopy (STS) experiments. 

To evaluate the DOS, we utilize the retarded normal Green's function $\hat{G}^R(\omega, \mathbf{k}, \phi)$, which corresponds to the diagonal electron-electron block of the full inverted Nambu-Gor'kov matrix~\cite{supplemental}. Performing an analytic continuation to the real frequency axis $i\omega_n \to \omega + i\Gamma$, the momentum-integrated DOS of the proximitized AM layer is given by:
\begin{equation} 
N(\omega) = -\frac{1}{\pi N_{k}} \sum_{\mathbf{k},\sigma} \text{Im}\left[ G_{\sigma\sigma}^{R}(\omega+i\Gamma, \mathbf{k}, \phi) \right],
\end{equation} 
where $N_k$ is the total number of momentum grid points in the Brillouin zone, and $\Gamma$ is the phenomenological Dynes parameter accounting for inelastic quasiparticle scattering rates and experimental broadening \cite{Dynes1978, Fischer2007}. We set $\Gamma=0.5 \Delta_d$ to capture the strong interfacial scattering in the proximitized AM layer~\cite{Takei2013}, as well as the intrinsic nodal quasiparticle lifetime typical of cuprates~\cite{Reber2012,Gu2019}.  

Crucially, while the energy bands are locally spin-split by the altermagnet, the total momentum-integrated DOS remains globally spin-symmetric: $N_\uparrow(\omega) = N_\downarrow(\omega)=N(\omega)/2$. This global symmetry is a direct consequence of the $d$-wave geometry. A $\pi/2$ rotation in momentum space maps the altermagnetic field to its negative ($h_{k_y, -k_x} = -h_{k_x, k_y}$). Because the proximitized self-energy depends strictly on the magnitude squared of the effective gap ($|\Delta_\phi(\mathbf{k})|^2$), the simultaneous sign change of the $d$-wave gap under this rotation leaves the Green's function denominator invariant. Consequently, summing over the entire Brillouin zone perfectly maps the spin-up Green's function onto its spin-down counterpart. We therefore focus on the total spin-summed DOS to capture the macroscopic spectroscopic signatures.

At half-filling ($\mu=0$), the normal state inherently harbors van Hove singularities due to band-structure critical points at the Brillouin zone axes, exactly where the $d_{x^2-y^2}$ altermagnetic field $h_{\mathbf{k}}$ is maximized. In the aligned configuration ($\phi = 0$, PME phase), the antinodes of the SC gap optimally overlap with these maximally spin-split critical points [Fig.~\ref{fig:ldos}(a)]. This strong local Pauli depairing suppresses static singlet pairing and drives a robust dynamical conversion into odd-frequency triplets. As a result, the DOS [Fig.~\ref{fig:ldos}(c)] exhibits sharp, prominent subgap resonance peaks near zero energy that broaden with increasing exchange strength $J$.

Conversely, at the maximal twist ($\phi = \pi/4$, Meissner phase), the SC antinodes rotate to the Brillouin zone diagonals [Fig.~\ref{fig:ldos}(b)]. These diagonals serve as magnetic ``safe havens" ($h_{\mathbf{k}}=0$) that protect the static singlet pairs and restore the robust diamagnetic state. Spectroscopically, the dominant odd-frequency resonances are entirely suppressed. The DOS [Fig.~\ref{fig:ldos}(d)] instead recovers a deep subgap profile bordered by conventional, albeit heavily damped, coherence peaks at the effective gap edges ($\omega \approx \pm 0.05t$). 
The residual zero-energy density simply reflects phenomenological impurity scattering ($\Gamma$) at the ungapped nodes rather than an anomalous triplet resonance. Thus, the twist angle effectively rotates the superconducting gap between magnetic ``kill zones" and ``safe havens," deterministically toggling both the subgap spectroscopic signatures and the macroscopic quantum phase.

\textit{Discussion and Conclusion---}We have theoretically demonstrated that twisted heterostructures of $d$-wave altermagnets and superconductors offer an unprecedented platform for generating and controlling odd-frequency superconductivity. The twist angle serves as a macroscopic switch, tuning the spatial overlap between the SC gap nodes and the AM spin-splitting nodes. This allows for driving the total superfluid density from a conventional diamagnetic state into an anomalous paramagnetic Meissner state, circumventing the need to alter intrinsic system parameters or employ destructive external magnetic fields. Beyond the paramagnetic Meissner effect, this twist-tunable odd-frequency state provides an ideal playground for exploring other anomalous macroscopic phenomena, such as spontaneous $\pi$-phase shifts in altermagnet-based Josephson junctions. 

Promising experimental platforms for realizing this physics involve heterostructures combining high-$T_c$ $d$-wave cuprate superconductors, such as exfoliated $\text{Bi}_2\text{Sr}_2\text{CaCu}_2\text{O}_{8+\delta}$ (Bi-2212) or $\text{YBa}_2\text{Cu}_3\text{O}_{7-\delta}$ (YBCO) \cite{Yu2019, Zhao2021}, with the rapidly expanding family of $d$-wave altermagnets. While thin films of prototypical bulk altermagnets, such as metallic $\text{RuO}_2$ \cite{Fedchenko2024} or $\text{KRu}_4\text{O}_8$, possess the requisite spin splitting, they are typically grown epitaxially and lack continuous angular tunability. Fortunately, recent breakthroughs have identified a plethora of 2D and van der Waals-compatible $d$-wave altermagnets perfectly suited for our proposed twistronic setup. Excellent candidates include the room-temperature metallic altermagnet $\text{KV}_2\text{Se}_2\text{O}$ \cite{Jiang2025}, transition-metal fluorides such as $\text{RuF}_4$ and $\text{VF}_4$ \cite{Sodequist2024}, and the vdW magnetic semiconductor $\text{CrPS}_4$ \cite{Cui2024}. Advances in mechanical tear-and-stack vdW fabrication techniques make the precise angular alignment required to probe this magnetic phase diagram highly feasible, paving the way for the next generation of superconducting spintronics and twistronic devices.


\textit{Data availability---}No experimental data were created in this study. The numerical data and codes used to generate the plots in this study are available from the corresponding author upon reasonable request.
\bibliography{refs.bib}

@article{Pixley_AnnRev2026,
   author = "Pixley, J. H. and Volkov, Pavel A.",
   title = "Twisted Nodal Superconductors", 
   journal= "Annual Review of Condensed Matter Physics",
   year = "2026",
   volume = "17",
   number = "Volume 17, 2026",
   pages = "183-205",
   doi = "https://doi.org/10.1146/annurev-conmatphys-031524-063257",
   url = "https://www.annualreviews.org/content/journals/10.1146/annurev-conmatphys-031524-063257",
   publisher = "Annual Reviews",
   issn = "1947-5462",
   type = "Journal Article"
  }

@article{Balatsky_PRB1992,
  title = {New class of singlet superconductors which break the time reversal and parity},
  author = {Balatsky, Alexander and Abrahams, Elihu},
  journal = {Phys. Rev. B},
  volume = {45},
  issue = {22},
  pages = {13125(R)--13128(R)},
  numpages = {0},
  year = {1992},
  month = {Jun},
  publisher = {American Physical Society},
  doi = {10.1103/PhysRevB.45.13125},
  url = {https://link.aps.org/doi/10.1103/PhysRevB.45.13125}
}

@article{Song_NatRevMat2025,
	author = {Song, Cheng and Bai, Hua and Zhou, Zhiyuan and Han, Lei and Reichlova, Helena and Dil, J. Hugo and Liu, Junwei and Chen, Xianzhe and Pan, Feng},
	date = {2025/06/01},
	doi = {10.1038/s41578-025-00779-1},
	id = {Song2025},
	isbn = {2058-8437},
	journal = {Nature Reviews Materials},
	number = {6},
	pages = {473--485},
	title = {Altermagnets as a new class of functional materials},
	url = {https://doi.org/10.1038/s41578-025-00779-1},
	volume = {10},
	year = {2025}}

@article{Ling_AdvFuncMat2024,
author = {Bai, Ling and Feng, Wanxiang and Liu, Siyuan and Šmejkal, Libor and Mokrousov, Yuriy and Yao, Yugui},
title = {Altermagnetism: Exploring New Frontiers in Magnetism and Spintronics},
journal = {Advanced Functional Materials},
volume = {34},
number = {49},
pages = {2409327},
doi = {https://doi.org/10.1002/adfm.202409327},
url = {https://advanced.onlinelibrary.wiley.com/doi/abs/10.1002/adfm.202409327},
year = {2024}
}

@misc{supplemental,
note={See the Supplemental Materials for details.}
}

@article{Fukaya_JPCM2025,
doi = {10.1088/1361-648X/adf1cf},
url = {https://doi.org/10.1088/1361-648X/adf1cf},
year = {2025},
month = {aug},
publisher = {IOP Publishing},
volume = {37},
number = {31},
pages = {313003},
author = {Fukaya, Yuri and Lu, Bo and Yada, Keiji and Tanaka, Yukio and Cayao, Jorge},
title = {Superconducting phenomena in systems with unconventional magnets},
journal = {Journal of Physics: Condensed Matter}
}

@article{
Smejkel_SciAdv2020,
author = {Libor {\v{S}}mejkal  and Rafael González-Hernández  and T. Jungwirth  and J. Sinova },
title = {Crystal time-reversal symmetry breaking and spontaneous {H}all effect in collinear antiferromagnets},
journal = {Science Advances},
volume = {6},
number = {23},
pages = {eaaz8809},
year = {2020},
doi = {10.1126/sciadv.aaz8809},
URL = {https://www.science.org/doi/abs/10.1126/sciadv.aaz8809}
}

@article{Jungwirth_Newton2025,
	author = {Jungwirth, Tom{\'a}{\v s} and Fernandes, Rafael M. and Fradkin, Eduardo and MacDonald, Allan H. and Sinova, Jairo and {\v S}mejkal, Libor},
	journal = {Newton},
	number = {6},
	title = {Altermagnetism: An unconventional spin-ordered phase of matter},
	volume = {1},
    pages={100162},
	year = {2025},
    doi = {10.1016/j.newton.2025.100162},
    }

@article{Zhao2021,
  title = {Sign-Reversing {H}all Effect in Atomically Thin High-Temperature {B}i$_{2.1}${Sr}$_{1.9}${C}a{C}u$_{2.0}${O}$_{8+\delta}$ Superconductors},
  author = {Zhao, S. Y. Frank and Poccia, Nicola and Panetta, Margaret G. and Yu, Cyndia and Johnson, Jedediah W. and Yoo, Hyobin and Zhong, Ruidan and Gu, G. D. and Watanabe, Kenji and Taniguchi, Takashi and Postolova, Svetlana V. and Vinokur, Valerii M. and Kim, Philip},
  journal = {Phys. Rev. Lett.},
  volume = {122},
  issue = {24},
  pages = {247001},
  numpages = {6},
  year = {2019},
  month = {Jun},
  publisher = {American Physical Society},
  doi = {10.1103/PhysRevLett.122.247001},
  url = {https://link.aps.org/doi/10.1103/PhysRevLett.122.247001}
}

@article{Smejkal2022,
  author = {L. {\v{S}}mejkal and J. Sinova and T. Jungwirth},
  title = {Emerging Research Landscape of Altermagnetism},
  journal = {Phys. Rev. X},
  volume = {12},
  pages = {040501},
  year = {2022},
  doi = {10.1103/PhysRevX.12.040501}
}

@article{Smejkal2022symmetry,
  author = {L. {\v{S}}mejkal and J. Sinova and T. Jungwirth},
  title = {Beyond Conventional Ferromagnetism and Antiferromagnetism: A Phase with Nonrelativistic Spin and Crystal Rotation Symmetry},
  journal = {Phys. Rev. X},
  volume = {12},
  pages = {031042},
  year = {2022},
  doi = {10.1103/PhysRevX.12.031042}
}

@article{Smejkal2022_PRX,
  author = {L. {\v{S}}mejkal and A. B. Hellenes and R. Gonz\'alez-Hern\'andez and J. Sinova and T. Jungwirth},
  title = {Giant and Tunneling Magnetoresistance in Unconventional Collinear Antiferromagnets},
  journal = {Phys. Rev. X},
  volume = {12},
  pages = {011028},
  year = {2022},
  doi = {10.1103/PhysRevX.12.011028}
}

@article{Mazin2022,
  title = {Editorial: Altermagnetism---A New Punch Line of Fundamental Magnetism},
  author = {Mazin, Igor},
  collaboration = {The PRX Editors},
  journal = {Phys. Rev. X},
  volume = {12},
  issue = {4},
  pages = {040002},
  numpages = {3},
  year = {2022},
  month = {Dec},
  publisher = {American Physical Society},
  doi = {10.1103/PhysRevX.12.040002},
  url = {https://link.aps.org/doi/10.1103/PhysRevX.12.040002}
}

@article{Ahn2019a,
  author = {K.-H. Ahn and A. Hariki and K.-W. Lee and J. Kune{\v{s}}},
  title = {Antiferromagnetism in {R}u{O}$_2$ as $d$-wave {P}omeranchuk instability},
  journal = {Phys. Rev. B},
  volume = {99},
  pages = {184432},
  year = {2019},
  doi = {10.1103/PhysRevB.99.184432}
}

@article{Ouassou2023,
  title = {dc {J}osephson Effect in Altermagnets},
  author = {Ouassou, Jabir Ali and Brataas, Arne and Linder, Jacob},
  journal = {Phys. Rev. Lett.},
  volume = {131},
  issue = {7},
  pages = {076003},
  numpages = {6},
  year = {2023},
  month = {Aug},
  publisher = {American Physical Society},
  doi = {10.1103/PhysRevLett.131.076003},
  url = {https://link.aps.org/doi/10.1103/PhysRevLett.131.076003}
}

@article{Berezinskii1974,
  author = {V. L. Berezinski\v{\i}},
  title = {New model of the anisotropic phase of superfluid {H}e$^3$},
  journal = {ZhETF Pis. Red.},
  volume = {20},
  pages = {628},
  year = {1974},
   note={[JETP Lett. \textbf{20}, 287 (1974)]}
}

@article{Linder2015,
  author = {J. Linder and A. V. Balatsky},
  title = {Odd-frequency superconductivity},
  journal = {Rev. Mod. Phys.},
  volume = {91},
  pages = {045005},
  year = {2019},
  doi = {10.1103/RevModPhys.91.045005}
}

@article{Tanaka2012,
  author = {Y. Tanaka and M. Sato and N. Nagaosa},
  title = {Symmetry and Topology in Superconductors --Odd-Frequency Pairing and Edge States--},
  journal = {J. Phys. Soc. Jpn.},
  volume = {81},
  pages = {011013},
  year = {2012},
  doi = {10.1143/JPSJ.81.011013}
}

@article{DiBernardo_PRX2015,
  title = {Intrinsic Paramagnetic {M}eissner Effect Due to $s$-Wave Odd-Frequency Superconductivity},
  author = {Di Bernardo, A. and Salman, Z. and Wang, X. L. and Amado, M. and Egilmez, M. and Flokstra, M. G. and Suter, A. and Lee, S. L. and Zhao, J. H. and Prokscha, T. and Morenzoni, E. and Blamire, M. G. and Linder, J. and Robinson, J. W. A.},
  journal = {Phys. Rev. X},
  volume = {5},
  issue = {4},
  pages = {041021},
  numpages = {7},
  year = {2015},
  month = {Nov},
  publisher = {American Physical Society},
  doi = {10.1103/PhysRevX.5.041021},
  url = {https://link.aps.org/doi/10.1103/PhysRevX.5.041021}
}

@article{DiBernardo_NatureComm2015,
	author = {Di Bernardo, A. and Diesch, S. and Gu, Y. and Linder, J. and Divitini, G. and Ducati, C. and Scheer, E. and Blamire, M. G. and Robinson, J. W. A.},
	date = {2015/09/02},
	doi = {10.1038/ncomms9053},
	id = {Di Bernardo2015},
	isbn = {2041-1723},
	journal = {Nature Communications},
	number = {1},
	pages = {8053},
	title = {Signature of magnetic-dependent gapless odd frequency states at superconductor/ferromagnet interfaces},
	url = {https://doi.org/10.1038/ncomms9053},
	volume = {6},
	year = {2015}
    }

@article{Asano2011,
  title = {Unconventional Surface Impedance of a Normal-Metal Film Covering a Spin-Triplet Superconductor Due to Odd-Frequency {C}ooper Pairs},
  author = {Asano, Yasuhiro and Golubov, Alexander A. and Fominov, Yakov V. and Tanaka, Yukio},
  journal = {Phys. Rev. Lett.},
  volume = {107},
  issue = {8},
  pages = {087001},
  numpages = {4},
  year = {2011},
  month = {Aug},
  publisher = {American Physical Society},
  doi = {10.1103/PhysRevLett.107.087001},
  url = {https://link.aps.org/doi/10.1103/PhysRevLett.107.087001}
}

@article{Fominov2015,
  title = {Odd-frequency superconducting states with different types of {M}eissner response: Problem of coexistence},
  author = {Fominov, Ya. V. and Tanaka, Y. and Asano, Y. and Eschrig, M.},
  journal = {Phys. Rev. B},
  volume = {91},
  issue = {14},
  pages = {144514},
  numpages = {14},
  year = {2015},
  month = {Apr},
  publisher = {American Physical Society},
  doi = {10.1103/PhysRevB.91.144514},
  url = {https://link.aps.org/doi/10.1103/PhysRevB.91.144514}
}

@article{Yokoyama2011,
  author = {T. Yokoyama and Y. Tanaka and N. Nagaosa},
  title = {Anomalous {M}eissner Effect in a Normal-Metal-Superconductor Junction with a Spin-Active Interface},
  journal = {Phys. Rev. Lett.},
  volume = {106},
  pages = {246601},
  year = {2011},
  doi = {10.1103/PhysRevLett.106.246601}
}

@article{Suzuki2014,
  title = {Paramagnetic instability of small topological superconductors},
  author = {Suzuki, Shu-Ichiro and Asano, Yasuhiro},
  journal = {Phys. Rev. B},
  volume = {89},
  issue = {18},
  pages = {184508},
  numpages = {7},
  year = {2014},
  month = {May},
  publisher = {American Physical Society},
  doi = {10.1103/PhysRevB.89.184508},
  url = {https://link.aps.org/doi/10.1103/PhysRevB.89.184508}
}

@article{Mironov2012,
  author = {S. V. Mironov and A. Mel'nikov and A. Buzdin},
  title = {Vanishing {M}eissner effect as a Hallmark of In-Plane {F}ulde-{F}errell-{L}arkin-{O}vchinnikov Instability in Superconductor-Ferromagnet Layered Systems},
  journal = {Phys. Rev. Lett.},
  volume = {109},
  pages = {237002},
  year = {2012},
  doi = {10.1103/PhysRevLett.109.237002}
}

@article{Alam2026,
  author = {O. Alam and A. Pal and P. Dutta and A. Saha},
  title = {Proximity-induced superconductivity and emerging topological phases in altermagnet-based heterostructures},
  journal = {Phys. Rev. B},
  volume = {113},
  pages = {155429},
  year = {2026},
  doi = {10.1103/PhysRevB.113.155429}
}

@article{Heinsdorf2026,
  author = {N. Heinsdorf and M. Franz},
  title = {Proximitizing altermagnets with conventional superconductors},
  journal = {Phys. Rev. B},
  volume = {113},
  pages = {L020501},
  year = {2026},
  doi = {10.1103/PhysRevB.113.L020501}
}

@article{Dynes1978,
  author = {R. C. Dynes and V. Narayanamurti and J. P. Garno},
  title = {Direct Measurement of Quasiparticle-Lifetime Broadening in a Strong-Coupled Superconductor},
  journal = {Phys. Rev. Lett.},
  volume = {41},
  pages = {1509},
  year = {1978},
  doi = {10.1103/PhysRevLett.41.1509}
}

@article{Fischer2007,
  author = {{\O}. Fischer and M. Kugler and I. Maggio-Aprile and C. Berthod and C. Renner},
  title = {Scanning tunneling spectroscopy of high-temperature superconductors},
  journal = {Rev. Mod. Phys.},
  volume = {79},
  pages = {353},
  year = {2007},
  doi = {10.1103/RevModPhys.79.353}
}

@article{Yu2019,
  author = {Y. Yu and L. Ma and P. Cai and R. Zhong and C. Ye and J. Shen and G. D. Gu and X. H. Chen and Y. Zhang},
  title = {High-temperature superconductivity in monolayer {B}i$_2${S}r$_2${C}a{C}u$_2${O}$_{8+\delta}$},
  journal = {Nature},
  volume = {575},
  pages = {156--163},
  year = {2019},
  doi = {10.1038/s41586-019-1718-x}
}

@article{Fedchenko2024,
author = {Olena Fedchenko  and Jan Minár  and Akashdeep Akashdeep  and Sunil Wilfred D’Souza  and Dmitry Vasilyev  and Olena Tkach  and Lukas Odenbreit  and Quynh Nguyen  and Dmytro Kutnyakhov  and Nils Wind  and Lukas Wenthaus  and Markus Scholz  and Kai Rossnagel  and Moritz Hoesch  and Martin Aeschlimann  and Benjamin Stadtmüller  and Mathias Kläui  and Gerd Schönhense  and Tomas Jungwirth  and Anna Birk Hellenes  and Gerhard Jakob  and Libor {\v{S}}mejkal  and Jairo Sinova  and Hans-Joachim Elmers },
title = {Observation of time-reversal symmetry breaking in the band structure of altermagnetic {R}u{O}$_2$},
journal = {Science Advances},
volume = {10},
number = {5},
pages = {eadj4883},
year = {2024},
doi = {10.1126/sciadv.adj4883},
URL = {https://www.science.org/doi/abs/10.1126/sciadv.adj4883},
}

@article{Jiang2025,
	author = {Jiang, Bei and Hu, Mingzhe and Bai, Jianli and Song, Ziyin and Mu, Chao and Qu, Gexing and Li, Wan and Zhu, Wenliang and Pi, Hanqi and Wei, Zhongxu and Sun, Yu-Jie and Huang, Yaobo and Zheng, Xiquan and Peng, Yingying and He, Lunhua and Li, Shiliang and Luo, Jianlin and Li, Zheng and Chen, Genfu and Li, Hang and Weng, Hongming and Qian, Tian},
	date = {2025/05/01},
	doi = {10.1038/s41567-025-02822-y},
	id = {Jiang2025},
	isbn = {1745-2481},
	journal = {Nature Physics},
	number = {5},
	pages = {754--759},
	title = {A metallic room-temperature $d$-wave altermagnet},
	url = {https://doi.org/10.1038/s41567-025-02822-y},
	volume = {21},
	year = {2025}
}

@article{Sodequist2024,
    author = {S{\o}dequist, Joachim and Olsen, Thomas},
    title = {Two-dimensional altermagnets from high throughput computational screening: Symmetry requirements, chiral magnons, and spin-orbit effects},
    journal = {Applied Physics Letters},
    volume = {124},
    number = {18},
    pages = {182409},
    year = {2024},
    month = {05},
    issn = {0003-6951},
    doi = {10.1063/5.0198285},
    url = {https://doi.org/10.1063/5.0198285}
}

@article{Cui2024,
    author = {Cui, Qirui and Bai, Xiaocheng and Ge, Yuqing and Edstr\"om, Alexander and Li, Cong and Sassa, Yasmine and Song, Cheng and Wang, Kaiyou and Delin, Anna},
    title = {Altermagnetic Magnons in Twisted van der {W}aals Antiferromagnets},
    journal = {Nano Letters},
    volume = {26},
    number = {15},
    pages = {5078-5085},
    year = {2026},
    month = {04},
    issn = {1530-6984},
    doi = {10.1021/acs.nanolett.6c00198},
    url = {https://doi.org/10.1021/acs.nanolett.6c00198}
}

@misc{Vakili2026,
     title={Supercurrent-Driven {N}{\'e}el Torque in Superconductor/Altermagnet Hybrids}, 
      author={Hamed Vakili and Moaz Ali and Igor Žutić and Alexey A. Kovalev},
      year={2026},
      eprint={2603.22243},
      archivePrefix={arXiv},
      primaryClass={cond-mat.mes-hall},
      url={https://arxiv.org/abs/2603.22243}, 
}

@article{Gu2019,
  author = {Gu, Q. and Wan, S. and Tang, Q. and Pan, Z. and Shi, Y. and Cai, P. and Song, Q. and Yin, Y. and Li, Y. and Zhao, Z. and Ding, H.},
  title = {Directly visualizing the sign change of $d$-wave superconducting gap in {B}i$_2${S}r$_2${C}a{C}u$_2${O}$_{8+\delta}$ by phase-referenced quasiparticle interference},
  journal = {Nature Communications},
  year = {2019},
  volume = {10},
  number = {1},
  pages = {1603},
  doi = {10.1038/s41467-019-09340-5}
}

@article{Takei2013,
  title = {Soft Superconducting Gap in Semiconductor {M}ajorana Nanowires},
  author = {Takei, So and Fregoso, Benjamin M. and Hui, Hoi-Yin and Lobos, Alejandro M. and Das Sarma, S.},
  journal = {Phys. Rev. Lett.},
  volume = {110},
  issue = {18},
  pages = {186803},
  numpages = {5},
  year = {2013},
  month = {Apr},
  publisher = {American Physical Society},
  doi = {10.1103/PhysRevLett.110.186803},
  url = {https://link.aps.org/doi/10.1103/PhysRevLett.110.186803}
}

@article{Reber2012,
  author = {Reber, T. J. and Plumb, N. C. and Sun, Z. and Zhao, Y. and Aiura, Y. and Murai, K. and McLeod, D. S. and Fluss, M. J. and Fujimori, S. and Balakirev, F. F. and Dai, P. and Barzykin, V. and Shen, Z. X. and Dessau, D. S.},
  title = {The origin and non-quasiparticle nature of {F}ermi arcs in {B}i$_2${S}r$_2${C}a{C}u$_2${O}$_{8+\delta}$},
  journal = {Nature Physics},
  year = {2012},
  volume = {8},
  pages = {606--610},
  doi = {10.1038/nphys2352}
}

@article{Fukaya2025,
  title = {Josephson effect and odd-frequency pairing in superconducting junctions with unconventional magnets},
  author = {Fukaya, Yuri and Maeda, Kazuki and Yada, Keiji and Cayao, Jorge and Tanaka, Yukio and Lu, Bo},
  journal = {Phys. Rev. B},
  volume = {111},
  issue = {6},
  pages = {064502},
  numpages = {15},
  year = {2025},
  month = {Feb},
  publisher = {American Physical Society},
  doi = {10.1103/PhysRevB.111.064502},
  url = {https://link.aps.org/doi/10.1103/PhysRevB.111.064502}
}

@article{Maeda_PRB2025,
  title = {Classification of pair symmetries in superconductors with unconventional magnetism},
  author = {Maeda, Kazuki and Fukaya, Yuri and Yada, Keiji and Lu, Bo and Tanaka, Yukio and Cayao, Jorge},
  journal = {Phys. Rev. B},
  volume = {111},
  issue = {14},
  pages = {144508},
  numpages = {14},
  year = {2025},
  month = {Apr},
  publisher = {American Physical Society},
  doi = {10.1103/PhysRevB.111.144508},
  url = {https://link.aps.org/doi/10.1103/PhysRevB.111.144508}
}

@article{Zhao_PRB2025,
  title = {Orientation-dependent transport in junctions formed by $d$-wave altermagnets and $d$-wave superconductors},
  author = {Zhao, Wenjun and Fukaya, Yuri and Burset, Pablo and Cayao, Jorge and Tanaka, Yukio and Lu, Bo},
  journal = {Phys. Rev. B},
  volume = {111},
  issue = {18},
  pages = {184515},
  numpages = {11},
  year = {2025},
  month = {May},
  publisher = {American Physical Society},
  doi = {10.1103/PhysRevB.111.184515},
  url = {https://link.aps.org/doi/10.1103/PhysRevB.111.184515}
}

@misc{liu_arxiv2025,
   title={Altermagnetism and Superconductivity: A Short Historical Review}, 
      author={Zhao Liu and Hui Hu and Xia-Ji Liu},
      year={2025},
      eprint={2510.09170},
      archivePrefix={arXiv},
      primaryClass={cond-mat.supr-con},
      url={https://arxiv.org/abs/2510.09170}, 
}

@article{Bergeret_RMP2005,
  title = {Odd triplet superconductivity and related phenomena in superconductor-ferromagnet structures},
  author = {Bergeret, F. S. and Volkov, A. F. and Efetov, K. B.},
  journal = {Rev. Mod. Phys.},
  volume = {77},
  issue = {4},
  pages = {1321--1373},
  numpages = {0},
  year = {2005},
  month = {Nov},
  publisher = {American Physical Society},
  doi = {10.1103/RevModPhys.77.1321},
  url = {https://link.aps.org/doi/10.1103/RevModPhys.77.1321}
}

@article{Chakraborty_PRB2025,
  title = {Constraints on superconducting pairing in altermagnets},
  author = {Chakraborty, Debmalya and Black-Schaffer, Annica M.},
  journal = {Phys. Rev. B},
  volume = {112},
  issue = {1},
  pages = {014516},
  numpages = {12},
  year = {2025},
  month = {Jul},
  publisher = {American Physical Society},
  doi = {10.1103/zylh-rqxl},
  url = {https://link.aps.org/doi/10.1103/zylh-rqxl}
}

\clearpage

\onecolumngrid
\section*{End Matter: Complementary Symmetries and Kinetic Alignment}

\begin{figure*} 
 \includegraphics[width = \linewidth]{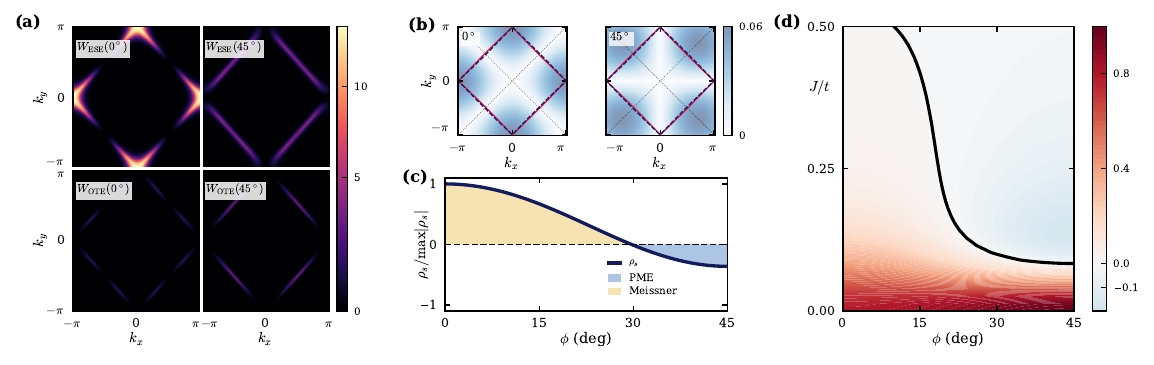} 
 \caption{Complementary magnetic phase switching in a twisted $d_{xy}$-wave AM/$d$-wave SC heterostructure. (a) Heatmaps of the $W_{\rm ESE}$ and $W_{\rm OTE}$ pairing weights at $\phi=0^\circ$ and $45^\circ$ for $J=0.1t$. (b) Fermi surface and momentum-space overlap of the $d_{xy}$ exchange field nodes (lines) with the SC gap (shading) for both the aligned ($\phi=0^\circ$) and maximally twisted ($\phi=45^\circ$) configurations. (c) Normalized $\rho_s$ vs. twist angle $\phi$. (d) The macroscopic $(J, \phi)$ phase diagram, revealing a highly robust PME pocket strictly centered around $\phi=45^\circ$ due to distinct kinetic alignment. Parameters used: $\mu=0$, $\lambda_s=0.15t$, $\Delta_d=0.03t$, and $k_{\rm B}T=0.02t$.}
 \label{fig:ldos_end1} 
\end{figure*}

\twocolumngrid

\textit{Phase Switching in $d_{xy}$-wave Altermagnets---}While the main text analyzed a $d_{x^2-y^2}$-wave altermagnet, our twist-angle mechanism applies universally to $d_{xy}$-wave altermagnets, albeit with complementary phase boundaries. For a $d_{xy}$ AM, the exchange field is rotated by $45^\circ$ ($h_\mathbf{k} \propto \sin k_x \sin k_y$). Consequently, the macroscopic magnetic response is completely inverted. As demonstrated in Fig.~\ref{fig:ldos_end1}, the system acts as a conventional diamagnet at perfect alignment ($\phi = 0^\circ$) due to the spatial shielding of the SC antinodes at the AM nodes, whereas it exhibits a robust paramagnetic Meissner state at maximal twist ($\phi = 45^\circ$).

\begin{figure}[b]
\centering
\includegraphics[width =\linewidth]{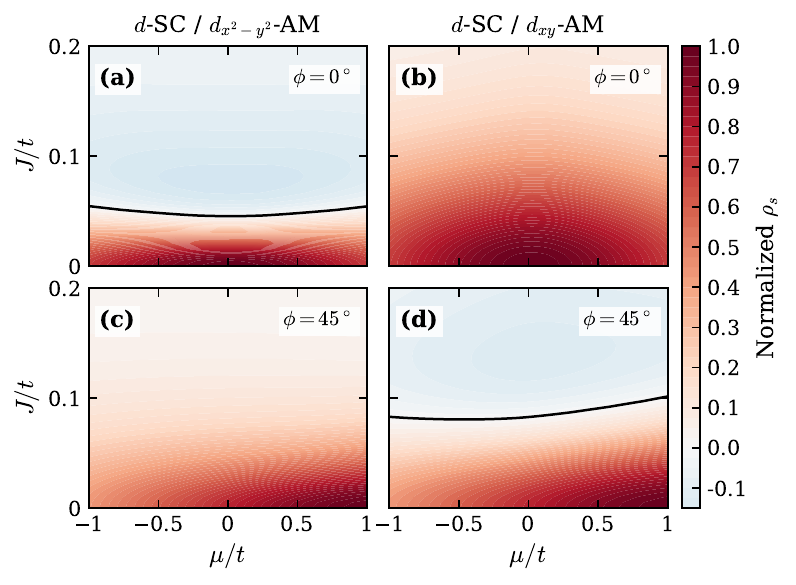} 
\caption{Magnetic phase diagram of $d$-wave AM/$d$-wave SC heterostructures in the parameter space of exchange strength $J$ and chemical potential $\mu$. (a),(c) A $d$-wave SC proximitized with a $d_{x^2-y^2}$ AM for the untwisted ($\phi=0^\circ$, top) and maximally twisted ($\phi=45^\circ$, bottom) structures. (b),(d) A $d$-wave SC proximitized with a $d_{xy}$ AM for the untwisted ($\phi=0^\circ$, top) and maximally twisted ($\phi=45^\circ$, bottom) structures. Other parameters used: $\lambda_s=0.15t$, $\Delta_d=0.03t$, and $k_{\rm B}T=0.02t$.}
\label{fig:phase_end} 
\end{figure}

\begin{figure*}
 \includegraphics[width = 0.8\linewidth]{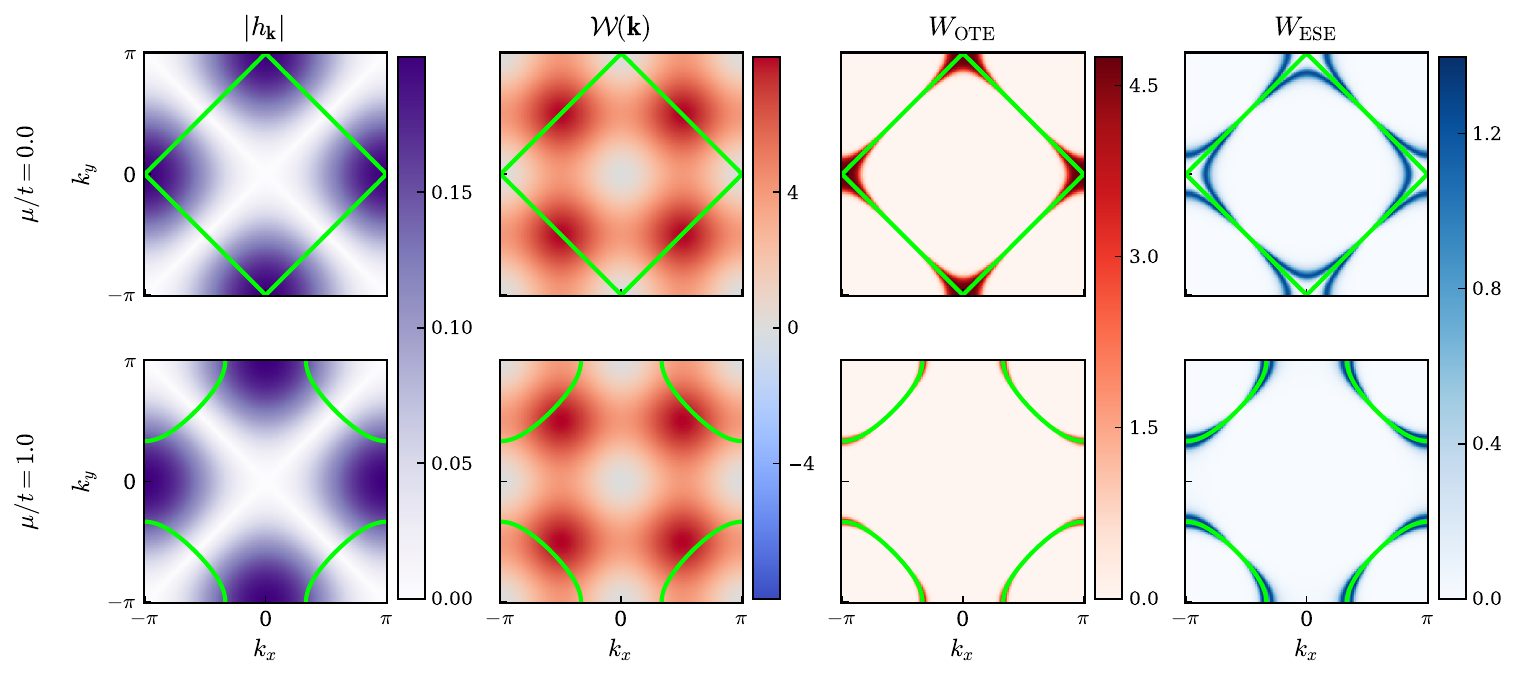} 
 \caption{Momentum-space evolution of pairing weights and Fermi surface topology for $J=0.05 t$. Heatmaps illustrating the spatial distribution of the altermagnetic exchange field magnitude $|h_{\mathbf{k}}|$, the kinetic weight $\mathcal{W}(\mathbf{k})$, the odd-frequency triplet weight $W_{\text{OTE}}$, and the even-frequency singlet weight $W_{\text{ESE}}$ for the aligned $d_{x^2-y^2}$-wave AM/$d$-wave SC heterostructure ($\phi=0^\circ$). The normal-state Fermi surface is superimposed as a solid green contour. Top row: At half-filling ($\mu=0$). Bottom row: At finite doping ($\mu=t$). The other model parameters are $\lambda_s=0.15 t$, $\Delta_d=0.03 t$ and $k_{\rm B}T=0.02 t$.}
 \label{fig:fermi_surface} 
\end{figure*}

Crucially, as shown in Fig.~\ref{fig:ldos_end1}(d), the $d_{xy}$ architecture unveils an exceptionally robust paramagnetic phase; a moderate increase in exchange strength $J$ strongly suppresses the singlet background, rendering the entire heterostructure paramagnetic across a vastly broadened angular range. The physical origin of this dominance lies in the momentum-space competition over the kinetic weight. At half-filling ($\mu \approx  0$), the maximum Fermi velocities (fast electrons) are strictly located along the Brillouin zone diagonals, while the axes host band-structure critical points with vanishing velocities. The $d_{xy}$ exchange field directly overlaps with these high-velocity diagonal states, heavily suppressing the kinetically active singlet pairs. In the absence of high-velocity singlet pairs, the dynamically generated odd-frequency triplets effortlessly dominate the macroscopic supercurrent, driving the global PME. In contrast, the $d_{x^2-y^2}$ exchange field primarily suppresses the zero-velocity axial states, leaving the fast diagonal singlets intact to protect the diamagnetic phase.

These contrasting kinetic overlaps directly manifest in the global $(J, \mu)$ magnetic phase diagrams (Fig.~\ref{fig:phase_end}). The phase boundaries confirm that the twist-tunable switching mechanism is remarkably universal but symmetry-dependent. While the $d_{x^2-y^2}$ AM/$d$-wave SC heterostructure (Fig.~\ref{fig:phase_end}, left column) achieves a peak paramagnetic response at $\phi=0^\circ$ and is diamagnetic at $\phi = 45^\circ$, replacing the altermagnet with a $d_{xy}$ symmetry yields perfectly complementary behavior (Fig.~\ref{fig:phase_end}, right column). Because the $d_{xy}$ nodes are intrinsically shifted by $45^\circ$ relative to the $d_{x^2-y^2}$ axes, the PME strictly emerges at the maximal twist angle ($\phi = 45^\circ$). This establishes the twist angle as a versatile, symmetry-selective tuning knob for controlling macroscopic quantum phases across a broad parameter space.

\textit{Role of Fermi Surface Tuning and Kinetic Weight---}To realize a robust macroscopic PME, generating odd-frequency triplets is necessary but not sufficient; these pairs must also carry a net supercurrent. The linear response superfluid density is weighted by the kinetic factor $\mathcal{W}(\mathbf{k}) =\sum_i(v_i^2 - u_i^2)$, where $v_i = \partial_{k_i} \xi_\mathbf{k} = 2t \sin k_i$ and $u_i = \partial_{k_i} h_\mathbf{k}$.

For the aligned $d_{x^2-y^2}$ architecture, the triplet antinodes are located at the Brillouin zone axes $(\pm\pi, 0)$ and $(0, \pm\pi)$. At half-filling ($\mu=0$), the Fermi surface crosses these exact points, which correspond to the saddle points of the normal state. Crucially, both the normal Fermi velocity ($v_x = v_y = 0$) and the altermagnetic velocity ($u_x = u_y = 0$) strictly vanish at these points, resulting in a zero kinetic weight ($\mathcal{W} = 0$). However, the diverging density of states in their immediate vicinity perfectly overlaps with the maximal exchange field $h_{\mathbf{k}}$. This allows the dynamically generated triplets to overwhelmingly dominate the local singlet background [Fig.~\ref{fig:fermi_surface} (top row)]. We note that while the altermagnet intrinsically lifts the spin degeneracy, the resulting spin-split Fermi surfaces appear as a single unresolved contour in these maps due to the small exchange magnitude $J=0.05t$. Tuning the chemical potential (e.g., $\mu=t$) shrinks the Fermi surface, effectively shifting the active integration region away from these magnetic hotspots. This spatial separation drastically suppresses the triplet weight $W_{\text{OTE}}$, while the singlet weight $W_{\text{ESE}}$ remains resilient, restoring the conventional diamagnetic Meissner state [Fig.~\ref{fig:fermi_surface} (bottom row)].

Similarly, for the $d_{xy}$ AM, tuning $\mu$ away from half-filling disconnects the Fermi surface from the highly mobile diagonal hotspots, gradually suppressing its PME. Furthermore, the structural twist ($\phi=45^\circ$) intrinsically breaks lattice commensurability, which manifests as a natural breaking of the exact particle-hole symmetry in the phase diagrams [Figs.~\ref{fig:phase_end}(c),(d)].

Crucially, this strong dependence of triplet mobility on the pairing hotspot locations stems fundamentally from the underlying square lattice rather than the superconducting gap anisotropy. This lattice-induced kinetic bias mathematically persists even if the $d$-wave SC is hypothetically replaced by an isotropic $s$-wave SC. An $s$-wave SC coupled to a $d_{x^2-y^2}$ AM remains strictly diamagnetic at half-filling because the triplet antinodes reside at the zero-velocity saddle points ($\mathcal{W}=0$). Conversely, coupling it to a $d_{xy}$ AM places the triplets exactly on the high-velocity diagonals ($\mathcal{W}=8t^2$), successfully mobilizing them to enable a PME at large exchange fields. True rotational invariance, where both AM symmetries yield identical macroscopic responses, would only emerge in a continuum limit (e.g., a two-dimensional electron gas), underscoring the critical role of crystal lattice symmetry in engineering macroscopic quantum phases.
\end{document}


\title{Supplementary Material for:// Twist-Tunable Paramagnetic Superconductivity in $d$-wave Altermagnet/Superconductor Heterostructures}

\author{Narges Kia}
\affiliation{Department of Physics, Institute for Advanced Studies in Basic Sciences (IASBS), Zanjan 45137-66731, Iran}

\author{Saeed H. Abedinpour}
\affiliation{Department of Physics, Institute for Advanced Studies in Basic Sciences (IASBS), Zanjan 45137-66731, Iran}

\author{Zahra Faraei}
\email{z.faraei@iasbs.ac.ir}
\affiliation{Department of Physics, Institute for Advanced Studies in Basic Sciences (IASBS), Zanjan 45137-66731, Iran}

\date{\today}

\maketitle

\setcounter{equation}{0}
\setcounter{figure}{0}
\setcounter{table}{0}
\setcounter{section}{0}
\renewcommand{\theequation}{S\arabic{equation}}
\renewcommand{\thefigure}{S\arabic{figure}}
\renewcommand{\thetable}{S\arabic{table}}
\renewcommand{\thesection}{S\Roman{section}} 

\section{MODEL HAMILTONIAN AND PROXIMITY EFFECT FORMALISM}

We consider a two-dimensional heterostructure composed of a single-layer $d$-wave altermagnet (AM) placed on top of a pristine $d$-wave superconductor (SC) substrate. The two layers are misaligned by a relative twist angle $\phi$. To describe the low-energy physics of this proximity-coupled system, we employ a Nambu-Gor'kov formalism. 

We define the basis in the Nambu spinor representation as $\Psi_{\mathbf{k}} = (c_{\mathbf{k}\uparrow}, c_{\mathbf{k}\downarrow}, c^\dagger_{-\mathbf{k}\downarrow}, -c^\dagger_{-\mathbf{k}\uparrow})^T$, where $c^\dagger_{\mathbf{k}\sigma}$ creates an electron with momentum $\mathbf{k}$ and spin $\sigma$ in the AM layer. The bare Hamiltonian of the AM layer in this basis is given by:
\begin{equation}
\mathcal{H}_{AM}(\mathbf{k}) = 
\begin{pmatrix} 
\xi_{\mathbf{k}} + h_{\mathbf{k}} & 0 & 0 & 0 \\ 
0 & \xi_{\mathbf{k}} - h_{\mathbf{k}} & 0 & 0 \\ 
0 & 0 & -\xi_{\mathbf{k}} + h_{\mathbf{k}} & 0 \\ 
0 & 0 & 0 & -\xi_{\mathbf{k}} - h_{\mathbf{k}} 
\end{pmatrix},
\end{equation}
which can be compactly written using Pauli matrices in particle-hole ($\tau_i$) and spin ($\sigma_i$) spaces as:
\begin{equation}
\mathcal{H}_{AM}(\mathbf{k}) = \xi_{\mathbf{k}} \tau_3 \sigma_0 + h_{\mathbf{k}} \tau_0 \sigma_3,
\end{equation}
where $\xi_{\mathbf{k}} = -2t(\cos k_x + \cos k_y) - \mu$ is the normal-state kinetic energy, and $h_{\mathbf{k}} = 2J(\cos k_x - \cos k_y)$ is the collinear $d_{x^2-y^2}$-wave spin-splitting exchange field characteristic of the altermagnet. Similarly, for $d_{xy}$-wave spin-splitting we have $h_{\mathbf{k}} = 2J\sin k_x \sin k_y$

The SC layer acts as an interacting reservoir. By treating the interlayer coupling via a phenomenological tunneling Hamiltonian and integrating out the SC degrees of freedom, the effect of the superconductor is imprinted onto the AM layer as a frequency- and momentum-dependent retarded self-energy $\hat{\Sigma}(i\omega_n, \mathbf{k}, \phi)$. 

In the Matsubara frequency domain [$\omega_n = (2n+1)\pi T$], assuming the wide-band limit for the bulk SC layer, the proximity-induced self-energy takes the form:
\begin{equation}\label{eq:self_energy}
\hat{\Sigma}(i\omega_n, \mathbf{k}, \phi) = -\lambda_s\frac{ i\omega_n \tau_0 \sigma_0 + \Delta_\phi(\mathbf{k}) \tau_1 \sigma_0}{\sqrt{\Delta_\phi^2(\mathbf{k}) + \omega_n^2}},
\end{equation}
where $\lambda_s$ is the effective proximity coupling strength. The function $\Delta_\phi(\mathbf{k})$ is the $d_{x^2-y^2}$-wave superconducting gap projected onto the AM Brillouin zone. Due to the relative twist angle $\phi$, the momentum coordinates are rotated, yielding:
\begin{equation}
\begin{aligned}
\Delta_\phi(\mathbf{k}) = \Delta_d \big[& \cos(k_x \cos\phi - k_y \sin\phi)\\
&- \cos(k_x \sin\phi + k_y \cos\phi) \big].    
\end{aligned}
\end{equation}

\section{EXACT DERIVATION OF THE ODD-FREQUENCY TRIPLET PAIRING}\label{GreenFunctionDerivation}

The effective dressed Green's function of the proximitized AM layer is obtained via the Dyson equation:
\begin{equation}
\mathcal{G}^{-1}(i\omega_n, \mathbf{k}, \phi) = \mathcal{G}_0^{-1}(i\omega_n, \mathbf{k}) - \hat{\Sigma}(i\omega_n, \mathbf{k}, \phi),
\end{equation}
where $\mathcal{G}_0^{-1} = i\omega_n \tau_0 \sigma_0 - \mathcal{H}_{AM}$. Substituting the self-energy from Eq.~\eqref{eq:self_energy}, we can write the full inverse Green's function as:
\begin{equation}
\mathcal{G}^{-1}(i\omega_n, \mathbf{k}, \phi) = i\tilde{\omega}_n \tau_0 \sigma_0 - \xi_{\mathbf{k}} \tau_3 \sigma_0 - h_{\mathbf{k}} \tau_0 \sigma_3 + \tilde{\Delta}_n(\mathbf{k}, \phi) \tau_1 \sigma_0.
\end{equation}
Here, we have introduced the renormalized Matsubara frequency and induced pairing gap:
\begin{equation}
\begin{aligned}
\tilde{\omega}_n(\mathbf{k}, \phi) &= \omega_n \left( 1 + \frac{\lambda_s}{\sqrt{\Delta_\phi^2(\mathbf{k}) + \omega_n^2}} \right), \\
\tilde{\Delta}_n(\mathbf{k}, \phi) &= \frac{\lambda_s \Delta_\phi(\mathbf{k})}{\sqrt{\Delta_\phi^2(\mathbf{k}) + \omega_n^2}}.
\end{aligned}
\end{equation}
To extract the full Green's functions, we invert the $4 \times 4$ matrix $\mathcal{G}^{-1}$. Due to the diagonal nature of the spin-dependent terms in our chosen basis, the matrix strictly decouples into two $2 \times 2$ blocks. Performing the exact algebraic inversion, the normal (diagonal) components of the Green's function are found to be:
\begin{align}
\mathcal{G}_{11}(i\omega_n, \mathbf{k}) &= \frac{i\tilde{\omega}_n + \xi_{\mathbf{k}} - h_{\mathbf{k}}}{(i\tilde{\omega}_n - h_{\mathbf{k}})^2 - \xi_{\mathbf{k}}^2 - \tilde{\Delta}_n^2(\mathbf{k}, \phi)}, \\
\mathcal{G}_{22}(i\omega_n, \mathbf{k}) &= \frac{i\tilde{\omega}_n + \xi_{\mathbf{k}} + h_{\mathbf{k}}}{(i\tilde{\omega}_n + h_{\mathbf{k}})^2 - \xi_{\mathbf{k}}^2 - \tilde{\Delta}_n^2(\mathbf{k}, \phi)}, \\
\mathcal{G}_{33}(i\omega_n, \mathbf{k}) &= \frac{i\tilde{\omega}_n - \xi_{\mathbf{k}} - h_{\mathbf{k}}}{(i\tilde{\omega}_n - h_{\mathbf{k}})^2 - \xi_{\mathbf{k}}^2 - \tilde{\Delta}_n^2(\mathbf{k}, \phi)}, \\
\mathcal{G}_{44}(i\omega_n, \mathbf{k}) &= \frac{i\tilde{\omega}_n - \xi_{\mathbf{k}} + h_{\mathbf{k}}}{(i\tilde{\omega}_n + h_{\mathbf{k}})^2 - \xi_{\mathbf{k}}^2 - \tilde{\Delta}_n^2(\mathbf{k}, \phi)}.
\end{align}
Similarly, the off-diagonal (anomalous) part of the Green's function matrix $\hat{\mathcal{F}}$ in spin space encodes the superconducting pairing amplitudes:
\begin{equation}
\hat{\mathcal{F}} = 
\begin{pmatrix} 
0 & \mathcal{F}_{\uparrow\downarrow} \\ 
\mathcal{F}_{\downarrow\uparrow} & 0 
\end{pmatrix}.
\end{equation}
To extract the physical pairing correlators from the full Green's function matrix, we must carefully consider the definition of the Matsubara Green's function, $\mathcal{G}_{ij} = -\langle \mathcal{T}_\tau \Psi_i \Psi^\dagger_j \rangle$, alongside our chosen Nambu basis $\Psi_{\mathbf{k}} = (c_{\mathbf{k}\uparrow}, c_{\mathbf{k}\downarrow}, c^\dagger_{-\mathbf{k}\downarrow}, -c^\dagger_{-\mathbf{k}\uparrow})^T$. For the spin-up-spin-down pairing, taking the 1st element of $\Psi_{\mathbf{k}}$ and the 3rd element of the adjoint spinor $\Psi^\dagger_{\mathbf{k}}$ yields $\mathcal{G}_{13} = -\langle c_{\mathbf{k}\uparrow} c_{-\mathbf{k}\downarrow} \rangle = -\mathcal{F}_{\uparrow\downarrow}$. Conversely, for the spin-down-spin-up pairing, the 4th element of the adjoint spinor contains an intrinsic minus sign ($\Psi^\dagger_{\mathbf{k}, 4} = -c_{-\mathbf{k}\uparrow}$). This internal minus sign exactly cancels the overall negative sign originating from the general Green's function definition, resulting in $\mathcal{G}_{24} = -\langle c_{\mathbf{k}\downarrow} (-c_{-\mathbf{k}\uparrow}) \rangle = +\langle c_{\mathbf{k}\downarrow} c_{-\mathbf{k}\uparrow} \rangle = \mathcal{F}_{\downarrow\uparrow}$.

Using these exact relations ($\mathcal{F}_{\uparrow\downarrow} = -\mathcal{G}_{13}$ and $\mathcal{F}_{\downarrow\uparrow} = \mathcal{G}_{24}$), the non-zero anomalous components are explicitly found to be:
\begin{equation}
\mathcal{F}_{\uparrow\downarrow}(i\omega_n, \mathbf{k}, \phi) = \frac{\tilde{\Delta}_n(\mathbf{k}, \phi)}{(i\tilde{\omega}_n - h_{\mathbf{k}})^2 - \xi_{\mathbf{k}}^2 - \tilde{\Delta}_n^2(\mathbf{k}, \phi)},
\end{equation}
\begin{equation}
\mathcal{F}_{\downarrow\uparrow}(i\omega_n, \mathbf{k}, \phi) = \frac{-\tilde{\Delta}_n(\mathbf{k}, \phi)}{(i\tilde{\omega}_n + h_{\mathbf{k}})^2 - \xi_{\mathbf{k}}^2 - \tilde{\Delta}_n^2(\mathbf{k}, \phi)}.
\end{equation}
The pairing amplitudes can be decomposed into spin-singlet ($\psi$) and collinear spin-triplet ($d_z$, with $S_z=0$) components using standard Pauli matrix projections:
\begin{equation}
\psi = \frac{1}{2} \left( \mathcal{F}_{\uparrow\downarrow} - \mathcal{F}_{\downarrow\uparrow} \right), \quad 
d_z = \frac{1}{2} \left( \mathcal{F}_{\uparrow\downarrow} + \mathcal{F}_{\downarrow\uparrow} \right).
\end{equation}
By substituting the explicit forms of $\mathcal{F}_{\uparrow\downarrow}$ and $\mathcal{F}_{\downarrow\uparrow}$ into these definitions, we obtain the exact analytical expressions for both the induced spin-singlet and spin-triplet amplitudes:
\begin{equation}
\psi(i\omega_n, \mathbf{k}, \phi) = \frac{\chi_n(\mathbf{k}, \phi) \tilde{\Delta}_n(\mathbf{k}, \phi)}{\chi_n^2(\mathbf{k}, \phi) + 4h_{\mathbf{k}}^2 \tilde{\omega}_n^2},
\label{eq:singlet}
\end{equation}
\begin{equation}
d_z(i\omega_n, \mathbf{k}, \phi) = \frac{2 i \tilde{\omega}_n h_{\mathbf{k}}  \tilde{\Delta}_n(\mathbf{k}, \phi)}{\chi_n^2(\mathbf{k}, \phi) + 4h_{\mathbf{k}}^2 \tilde{\omega}_n^2},
\label{eq:triplet}
\end{equation}
where we have defined the characteristic denominator $\chi_n(\mathbf{k}, \phi) = h_{\mathbf{k}}^2 - \xi_{\mathbf{k}}^2 - \tilde{\omega}_n^2 - \tilde{\Delta}_n^2(\mathbf{k}, \phi)$. 

The contrasting symmetries of these two pairing states are immediately apparent from their numerators. Because $\chi_n$ and $\tilde{\Delta}_n$ are strictly \textit{even} functions of Matsubara frequency, the singlet component mathematically satisfies $\psi(i\omega_n) = \psi(-i\omega_n)$, placing it in the conventional even-frequency class. Conversely, the presence of $i\tilde{\omega}_n$ in the numerator of $d_z$ guarantees $d_z(i\omega_n) = -d_z(-i\omega_n)$. This mathematically proves that the dynamically generated triplet pairing inherently belongs to the exotic odd-frequency, spin-triplet, even-parity (OTE) symmetry class.


\section{LINEAR RESPONSE THEORY AND SUPERFLUID DENSITY}
To derive the macroscopic electromagnetic response, we evaluate the superfluid density $\rho_{s}$ using the linear response Kubo formalism in the Nambu-Gor'kov representation.

We introduce a static external vector potential $\mathbf{A}$. The coupling of the electromagnetic field to the system is implemented via the Peierls substitution $\mathbf{k} \rightarrow \mathbf{k} - e\mathbf{A}$ in the normal-state Hamiltonian $\mathcal{H}_{AM}(\mathbf{k}) = \xi_{\mathbf{k}} \tau_3 \sigma_0 + h_{\mathbf{k}} \tau_0 \sigma_3$. Expanding the Hamiltonian to second order in $\mathbf{A}$ yields the paramagnetic ($\hat{j}^p_i$) and diamagnetic ($\hat{j}^d_i$) current operators:
\begin{equation}
\hat{j}^{p}_i = -e \sum_{\mathbf{k}} \Psi_{\mathbf{k}}^\dagger \left( v_i \tau_3 \sigma_0 + u_i \tau_0 \sigma_3 \right) \Psi_{\mathbf{k}},
\end{equation}
\begin{equation}
\hat{j}^{d}_i = -e^2  A_i \sum_{\mathbf{k}} \Psi_{\mathbf{k}}^\dagger \left(\partial_{k_i} v_i \tau_3 \sigma_0 + \partial_{k_i} u_i \tau_0 \sigma_3 \right) \Psi_{\mathbf{k}},
\end{equation}
where $v_i = \partial_{k_i} \xi_{\mathbf{k}}$ is the standard normal-state Fermi velocity and $u_i = \partial_{k_i} h_{\mathbf{k}}$ is the spin-dependent altermagnetic velocity.

According to linear response theory, the total electromagnetic response kernel $K_{ii}$ is the sum of the diamagnetic and paramagnetic contributions ($K_{ii}=K^{dia}_{ii}+K^{para}_{ii}$). The diamagnetic kernel is determined by the expectation value of $\hat{j}^d_i$: 
\begin{equation}
K^{dia}_{ii} = e^2 T \sum_{\omega_n} \int_{\text{BZ}} \frac{d^2k}{(2\pi)^2} \text{Tr} \left[ \left( \partial_{k_i} v_i \tau_3 \sigma_0 + \partial_{k_i} u_i \tau_0 \sigma_3 \right) \mathcal{G} \right].
\end{equation}
The paramagnetic kernel is obtained from the static limit ($q\rightarrow 0$, $i\Omega_n\rightarrow 0$) of the current-current correlation function (bubble diagram):
\begin{equation}
K^{para}_{ii} = - e^2 T \sum_{\omega_n} \int_{\text{BZ}} \frac{d^2k}{(2\pi)^2} \text{Tr} \left[ \hat{J}_i \mathcal{G}(i\omega_n,\mathbf{k}) \hat{J}_i \mathcal{G}(i\omega_n,\mathbf{k}) \right],
\end{equation}
where the vertex matrix is defined as $\hat{J}_i = v_i \tau_3 \sigma_0 + u_i \tau_0 \sigma_3$.

To isolate the superconducting response, we separate the Green's function into its normal (single particle, diagonal) and anomalous (pairing, off-diagonal) components. For a normal metal, integration by parts demonstrates that the normal-state paramagnetic kernel exactly cancels the diamagnetic kernel ($K^{dia}_{ii}+K^{para}_{ii,\text{normal}}=0$), satisfying the standard diamagnetic sum rule. Consequently, the superfluid density $\rho_s$, which is proportional to the total static response, is entirely determined by the anomalous part of the paramagnetic kernel. 

Evaluating the algebraic trace over the $4 \times 4$ matrices derived in section~\ref{GreenFunctionDerivation} yields the anomalous contribution. The vertex matrix is strictly diagonal: $\hat{J}_i = \text{diag}(v_i+u_i, v_i-u_i, -v_i+u_i, -v_i-u_i)$. The trace reduces to products involving the off-diagonal anomalous Green's functions $\mathcal{G}_{13} = -(\psi + d_z)$ and $\mathcal{G}_{24} = -(\psi - d_z)$:
\begin{equation}
\text{Tr}[\hat{J}_i \mathcal{G} \hat{J}_i \mathcal{G}]_{\text{anom}} = -2(v_i^2 - u_i^2) \left[ \mathcal{G}_{13}^2 + \mathcal{G}_{24}^2 \right].
\end{equation}
Substituting the definitions of the pairing amplitudes, the cross-terms algebraically cancel out: $\mathcal{G}_{13}^2 + \mathcal{G}_{24}^2 = 2(\psi^2 + d_z^2)$. 
The total normalized superfluid density is thus given by summing the response over the spatial directions $i \in \{x,y\}$:
\begin{equation}
\rho_{s}(\phi) \propto T \sum_{\omega_n} \int_{\text{BZ}} d^2k \ \mathcal{W}(\mathbf{k}) \left[ \psi^2 + d_z^2 \right],
\end{equation}
where the exact effective kinetic weight factor is:
\begin{equation}
\mathcal{W}(\mathbf{k}) = \sum_{i=x,y}\left[(\partial_{k_i} \xi_{\mathbf{k}})^2 - (\partial_{k_i} h_{\mathbf{k}})^2\right].
\end{equation}
The relative sign difference between the singlet and triplet contributions in the superfluid density arises fundamentally from their respective symmetries in Matsubara space. The spin-singlet amplitude $\psi(i\omega_n)$ is an even, purely real function of frequency, meaning $\psi^2=|\psi|^2$. Conversely, the odd-frequency spin-triplet amplitude $d_z \propto i\omega_n$ is purely imaginary, which strictly dictates that $d_z^2=-|d_z|^2$. Thus, the macroscopic response mathematically reduces to:
\begin{equation}
\rho_{s}(\phi) \propto T \sum_{\omega_n} \int_{\text{BZ}} d^2k \ \mathcal{W}(\mathbf{k}) \left[ |\psi|^2 - |d_z|^2 \right].
\end{equation}
The negative sign preceding $|d_z|^2$ indicates that the odd-frequency pairing yields a paramagnetic supercurrent, i.e., the Paramagnetic Meissner Effect.

For the $d_{x^2-y^2}$ model considered in the main text, $\xi_{\mathbf{k}} = -2t(\cos k_x + \cos k_y) - \mu$ and $h_{\mathbf{k}} = 2J(\cos k_x - \cos k_y)$. Evaluating the derivatives yields $v_i = 2t \sin k_i$, and the spin-dependent altermagnetic velocities are $u_x = -2 J \sin k_x$ and $u_y = 2 J \sin k_y$. The total exact two-dimensional kinetic weight evaluates to:
\begin{equation}
\mathcal{W}_{x^2-y^2}(\mathbf{k}) = 4 (t^2 - J^2) \left( \sin^2 k_x + \sin^2 k_y \right).
\end{equation}

Alternatively, for a $d_{xy}$-wave altermagnetic model, the exchange field is defined as $h_{\mathbf{k}} = 2J \sin k_x \sin k_y$. Evaluating the momentum derivatives yields the spin-dependent altermagnetic velocities $u_x = 2J \cos k_x \sin k_y$ and $u_y = 2J \sin k_x \cos k_y$. Substituting these into the exact two-dimensional kinetic weight expression, the macroscopic response weight for the $d_{xy}$ architecture evaluates to:
\begin{equation}
\begin{aligned}
\mathcal{W}_{xy}(\mathbf{k}) =& \ 4t^2 \left( \sin^2 k_x + \sin^2 k_y \right) \\
&- 4J^2 \left( \cos^2 k_x \sin^2 k_y + \sin^2 k_x \cos^2 k_y \right).
\end{aligned}
\end{equation}